\documentclass[lettersize,journal]{IEEEtran}
\usepackage{amsmath,amsfonts}
\usepackage{algorithmic}
\usepackage{algorithm}
\usepackage{array}
\usepackage[caption=false,font=normalsize,labelfont=sf,textfont=sf]{subfig}
\usepackage{textcomp}
\usepackage{stfloats}
\usepackage{url}
\usepackage{verbatim}
\usepackage{graphicx}
\usepackage{cite}
\usepackage{multirow}
\usepackage{tabularx}
\usepackage{booktabs}
\usepackage{adjustbox}
\usepackage[none]{hyphenat}

\begin{document}

\title{Concept-based Explainable Malignancy Scoring on Pulmonary Nodules in CT Images}

\author{
    \IEEEauthorblockN{
        Rinat I. Dumaev\IEEEauthorrefmark{1},
        Sergei A. Molodyakov\IEEEauthorrefmark{2} and 
        Lev V. Utkin\IEEEauthorrefmark{3}
    }
    
    \IEEEauthorblockA{
    Peter the Great St.Petersburg Polytechnic University, St.Petersburg, Russia
        \\
        e-mail: \IEEEauthorrefmark{1}dumaevrinat@gmail.com,
        \IEEEauthorrefmark{2}samolodyakov@mail.ru,
        \IEEEauthorrefmark{3}lev.utkin@gmail.com
    }
}

\maketitle

\begin{abstract}

To increase the transparency of modern computer-aided diagnosis (CAD) systems for assessing the malignancy of lung nodules, an interpretable model based on applying the generalized additive models and the concept-based learning is proposed. The model detects a set of clinically significant attributes in addition to the final malignancy regression score and learns the association between the lung nodule attributes and a final diagnosis decision as well as their contributions into the decision. The proposed concept-based learning framework provides human-readable explanations in terms of different concepts (numerical and categorical), their values, and their contribution to the final prediction.

Numerical experiments with the LIDC-IDRI dataset demonstrate that the diagnosis results obtained using the proposed model, which explicitly explores internal relationships, are in line with similar patterns observed in clinical practice. Additionally, the proposed model shows the competitive classification and the nodule attribute scoring performance, highlighting its potential for effective decision-making in the lung nodule diagnosis.

\end{abstract}


\section{Introduction}

In current medical image analysis research, deep learning models have come to preeminence. The implementation of these advanced machine learning architectures has led to significant enhancements in diagnostic precision for various medical imaging challenges. In certain instances, the generated outcomes have even exceeded the capabilities of human experts \cite{wang2021review,xu2021deep,bonavita2020integration,majkowska2020chest}. Although these models have made significant progress, they are not widely used in real-world medical settings. The reason for this is that clinicians want to understand how the algorithm or model arrives at its decisions before they can rely on it.

To address the challenge of understanding the decision-making process of deep learning models, several approaches have been proposed that utilize saliency maps to highlight the contributions of individual regions or pixels to the model's predictions. However, recent studies have demonstrated that these strategies are unreliable and subjective. In addition, these models lack meaningfulness when applied to incorrect predictions \cite{Rudin2019, adebayo2018sanity}. In \cite{Rudin2019}, authors argue against the use of black box machine learning models for critical decision-making processes. Instead, it advocates for the development of inherently interpretable models that are designed from the outset to be transparent and understandable by humans. These models provide their own explanations that align with what the model computes, ensuring clarity and accountability in decision-making processes. Interpretable models are constrained in their form to be useful or to adhere to domain-specific knowledge, such as sparsity or causality, making them more suitable for high-stakes applications like healthcare.

Explainable AI techniques for images come in various forms and offer explanations through different modalities, including textual explanations \cite{hendricks2018grounding, zhang2017mdnet},  quantitative measures illustrating the importance of abstract concepts \cite{kim2018interpretability, chen2019looks} and feature-relevance visualizations \cite{ribeiro2016should, fong2019understanding, wang2020score, chen2019looks}.

According to the radiological lung nodule diagnosis guideline \cite{truong2014update}, clinicians are advised to carefully evaluate various morphological features, such as margin smoothness and spiculation contour, when making a diagnosis. Clinicians tend to communicate with each other using abstract concepts, such as "spiculation", "lobulation" and etc. to describe the particular severity of a condition, commonly used to describe the phenotype of a pulmonary nodule in a radiology report. This way of thinking is not unique to clinicians, as people in general tend to understand complex ideas through high-level concepts that are easily interpretable and translatable within a specific domain. Concept-based approaches are increasingly popular in medical imaging analysis due to their ability to break down the decision-making process into understandable human concepts \cite{chen2020concept, fang2020concept, graziani2018regression, lucieri2020interpretability, meldo2020natural}. These methods activate specific filters in the network when a concept is present in an image, allowing for both concept identification and class prediction. During inference, the concepts identified by the network in the original image can be determined.

In this study, we aim to develop a new scheme that can profile pulmonary nodules from a CT volumetric image with a concept-attribute vector. Instead of singleton malignancy scoring, this new scheme can yield several quantitative concept scores with reference to the concept features that may be more referential for clinical usage. We propose a CNN-based deep multi-task learning model to simultaneously score multiple attributes in lung nodules. On the first stage of the model, a deep CNN-based shared feature extractor encodes input volumetric image of nodule to feature embedding in a low-dimensional latent space. 

To calculate the contribution values of multiple numeric concepts and categorical features and predict the final malignancy regression score of a lung nodule, we propose generalized additive models (GAMs) as a decision explainer that are designed to be inherently interpretable. This part learns a linear combination of neural networks, each focusing on a single input concept, allowing them to capture complex relationships between features and outputs. GAMs and Neural Additive Models (NAMs) \cite{agarwal2021neural} have been shown to be more accurate than traditional interpretable models such as logistic regression and shallow decision trees \cite{agarwal2021neural, yang2021gami}. Interpreting NAMs is easy, as the impact of a feature on the prediction does not depend on the other features.

Numerical experiments with the LIDC-IDRI dataset demonstrate that the diagnosis results obtained using the proposed model, which explicitly explores internal relationships, are in line with similar patterns observed in clinical practice.

\begin{figure*}
\centering
\includegraphics[width=1\linewidth]{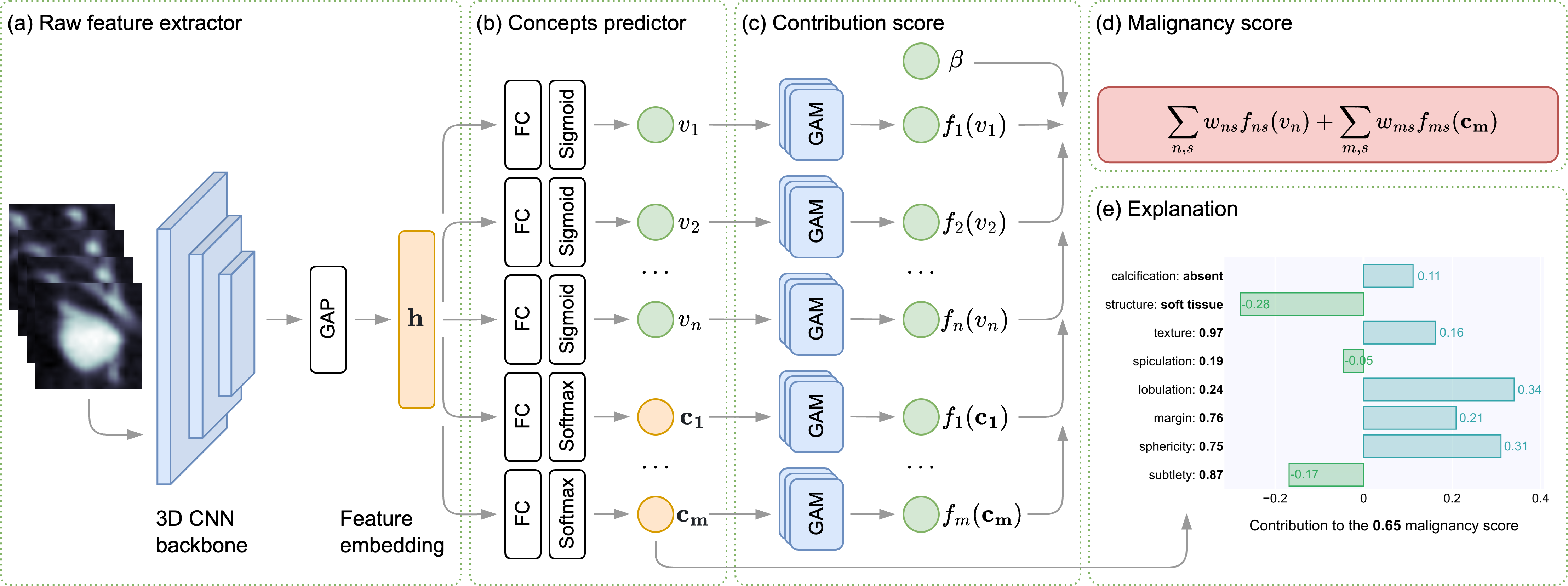}
\caption{\textbf{Overview of the proposed framework.} Our approach is divided into several components. (a) Deep 3D CNN-based shared feature extractor encodes input
image slices of the nodule to feature embedding in a low-dimensional latent space. (b) Learning numerical (e.g. subtlety, lobulation, margin) and categorical concept features (e.g. internal structure and calcification) by raw feature embedding. (c, d) Multiple GAM subnetworks are trained on each concept input feature and weighted sums are learned over the subnetworks. The outputs corresponding to each concept are summed and a bias is added to obtain the final malignancy score. (e) Predicted numerical and categorical concept values and corresponding calculated contributions are used to explain the predicted malignancy score. }
\label{fig:architecture}
\end{figure*}

\section{Related work}

\subsection{Concept-Based Models}

Concept-based learning approaches rely on the idea of utilizing human-defined concepts as an intermediate step to arrive at the final predictions. CBMs is an encoder-decoder architecture, where the encoder predicts the underlying concepts of the input image, and the decoder uses these concepts to generate the final output. This approach allows for the integration of domain knowledge and human expertise into the model. 

The concept of utilizing concept bottlenecks in deep neural networks has been explored in various studies, initially in the context of few-shot learning approaches \cite{lampert2009learning, kumar2009attribute}. More recently, the idea with the development of Concept Bottleneck Models (CBM) has been revived and applied to a wide range of tasks \cite{chen2020concept, kazhdan2020now, koh2020concept, wickramanayake2021comprehensible}. Since CBMs does not explain how a certain concept affects the final prediction, we introduce generalized additive models as a decoder part of CBM.

\subsection{Interpretable Malignancy Scoring on Pulmonary Nodules}

More recent methods are based on deep learning using CNN, and multi-task learning has become the preferred approach for simultaneously evaluating multiple attributes of lung nodules \cite{chen2016automatic, liu2019multi, dai2018incorporating, dumaev2023classification}.

Previous research has primarily focused on the indirect relationship between malignancy and other attributes. Malignancy or benignity can be inferred from certain concepts, such as the pattern of calcification (central, diffuse, popcorn and laminated), the well-defined nodule margin of the nodule or the internal composition of the nodule \cite{ost2012decision, macmahon2017guidelines, cruickshank2019evaluation, truong2014update}. Previous studies have employed a two-step approach, first predicted these characteristics and then used intermediate CNN features to infer malignancy without calculation of contribution values of concepts to the final malignancy regression score \cite{shen2019interpretable, mehta2021lung, wu2018joint}.

\section{Methods}

We propose a model that is not only capable of making decisions but also explaining the reasoning behind those decisions using human-comprehensible concepts. The interpretability of our model is achieved by estimating the contribution of each concept towards the target decision.
To accomplish this, we have incorporated a concept predictor component on top of a traditional shared feature extractor, as shown in Fig.~\ref{fig:architecture}. The role of the concept extractor is to identify the presence of a particular concept within a volumetric image or predict the score of a numerical concept. 

Furthermore, we added Generalized Additive sub-models to our framework, which enables us to disentangle the decision-making process and better understand the relationship between different concepts and the final decision. By using GAMs, we can decompose the decision-making process into additive components, making it easier to interpret and explain the model's predictions. 

In the following sections, we describe the different components of the proposed framework.

\subsection{Framework Architecture}

To formalize our problem, consider a training set $S=\{(\mathbf{x}_i, \mathbf{y}_i)\}^N_{i=1}$ where N is the number of training samples, $\mathbf{x}_i$ is an input image and $\mathbf{y}_i$ is the corresponding ground truth. We consider the raw feature extractor $\phi$ is a CNN backbone with global average pooling layer to reduce the dimensionality by averaging over each feature channel which takes $\mathbf{x}_i$ and maps into latent embedding $\mathbf{h}_i$. The feature embedding is then integrated into the concept predictor $\psi$ that maps $\mathbf{h}_i$ into human-interpretable concepts $\mathbf{c}_i$ and $\mathbf{v}_i$, where $\mathbf{v}_i = \{v^1_i, ..., v^n_i\}$ is the output of $n$ regression concept scores and $\mathbf{c}_i = \{c^1_i, ..., c^m_i\}$ is the output of $m$ categorical concepts. We consider the concept predictor $\psi$ that is a typical general multi-task learning model, which branches into task-specific pathways and becomes more specialized for individual concept attributes.

To calculate the contribution score multiple GAM subnetworks $\mathbf{f}_i$ are trained on each concept input feature $\mathbf{c}_i$ and $\mathbf{v}_i$, and weighted sums are learned over the subnetworks. The outputs corresponding to each concept are summed and a bias is added to obtain the final malignancy score.

\subsection{Shared Feature Extractor}

The shared feature extractor $\phi$ is composed with deep 3D CNN model (for which we used several architectures, described in detail in Section~\ref{section:experiments}) and global average pooling layer (GAP) to vectorize the feature volumes result after CNN model into the raw feature vector $\mathbf{h}_i$, reducing the dimensionality by averaging over each feature channel. The shared feature extractor learns to identify common visual attributes that will be used in the downstream concept predictor.

\subsection{Concepts Predictor}

The concepts predictor $\psi$ maps the CNN-extracted raw feature embedding $\mathbf{h}_i$ to regression concept scores $\mathbf{c}_i$ and categorical concepts $\mathbf{v}_i$. This module branches into task-specific pathways and contains fully-connected layers with the Rectified Linear Unit (ReLU) activation function and the final $sigmoid$ activation function for regression concept tasks (e.g. texture, spiculation) due to each concept score $v_m \in [0, 1]$  and the final $softmax$ activation function for categorical concept tasks (calcification and internal structure).

To train the concept predictor, we minimize the MSE loss between predicted and ground truth numerical concept scores and the final malignancy score:

\begin{equation}
    \mathcal{L}_{MSE} = \frac{1}{N}\sum_{i=1}^N(y_i - \hat{y}_i)^2.
\end{equation}

For categorical concepts, we use categorical cross-entropy loss:

\begin{equation}
    \mathcal{L}_{CE} = -\sum_{i=1}^Cy_i\log(\hat{y}_i),
\end{equation} where $C$ is the number of categorical concept classes and $\hat{y}$ is the predicted value.

\subsection{Decision Explainer}

\begin{figure}
\centering
\includegraphics[width=1\linewidth]{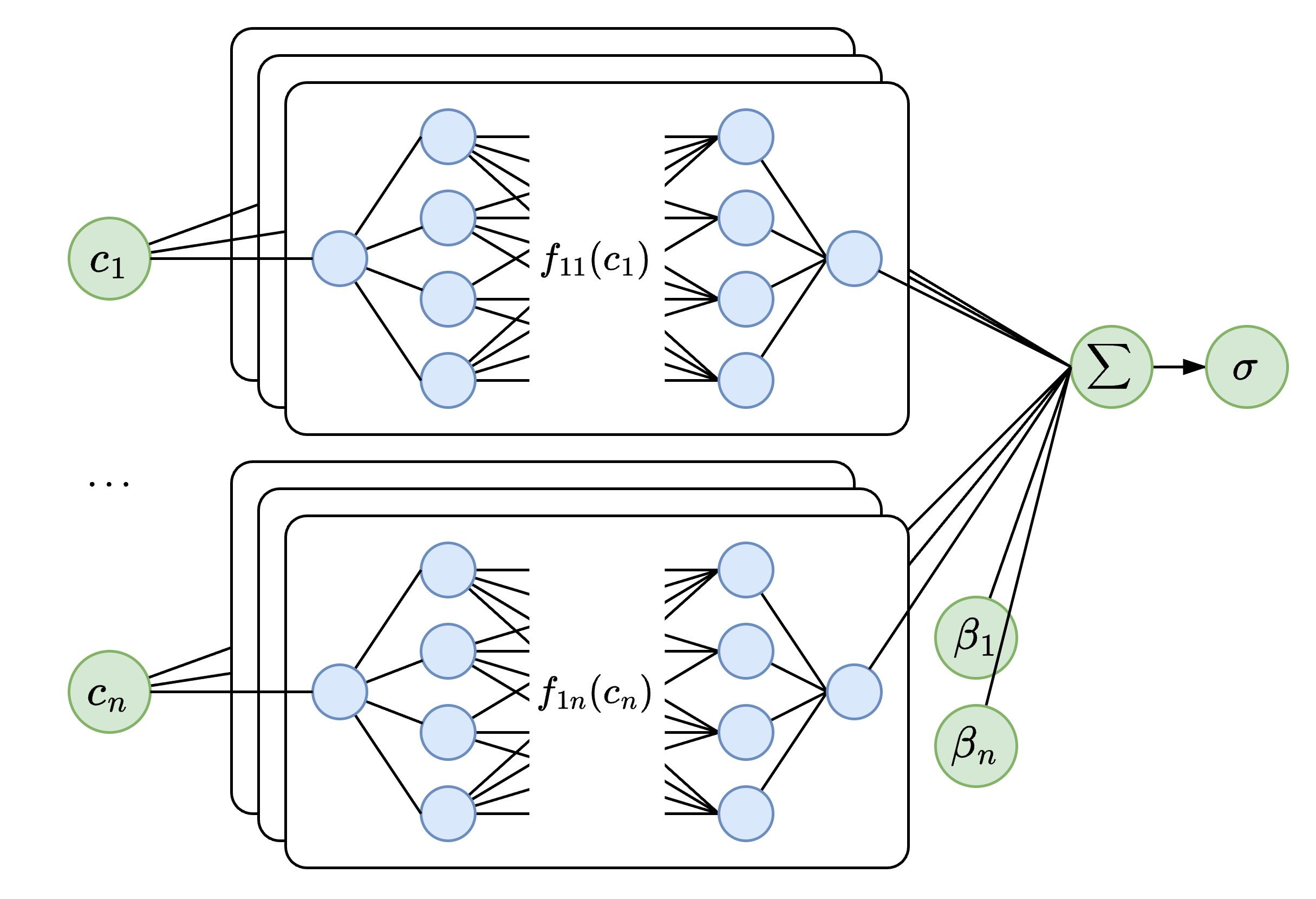}
\caption{\textbf{GAM architecture.} Multiple sub networks are trained on each input concept and
weighted sums are learned over the sub networks}
\label{fig:gam_architecture}
\end{figure}

To disentangle the decision-making process and better understand the relationship between different concept scores and predicted final malignancy score we use GAMs as a decision explainer. Fig. \ref{fig:gam_architecture} shows a multitask GAM architecture that can jointly learn different concept representations. By using GAMs, we can decompose the decision-making process into additive components \cite{agarwal2021neural}. In this component, each concept of $\mathbf{v}_n$ and $\mathbf{c}_m$ is associated with multiple subnetworks, and weighted sums are learned over the subnetworks within one concept. The outputs corresponding to each task are summed, and a bias is added to obtain the final malignancy score. The final malignancy score can be defined as follows:

\begin{equation}
    Malignancy_i = \sum_{n,s}w_{ns}f_{ns}(v^n_i) + \sum_{m,s}w_{ms}f_{ms}(\mathbf{c^m_i}) + \beta,
\end{equation} where $f_{ns}(v^n_i)$ and $f_{ms}(\mathbf{c^m_i})$ are shape functions for each sub-network in GAM and for each concept, $w_{ns}$ and $w_{ms}$ are weights related to each GAM sub-network for $n$ numerical concepts and $m$ categorical concepts. Outputs of shape functions corresponding to each concept can be interpreted as the impact of one concept on the prediction, which does not depend on the other concepts.

Taking into consideration all the above-described losses, the final loss is defined as:

\begin{equation}
    \mathcal{L} = \mathcal{L}_{MSE}(y, \hat{y}) + \mathcal{L}_{CE}(y, \hat{y}).
\end{equation}

\section{Experiments}
\label{section:experiments}

\subsection{Experimental Setup}

\subsubsection{Dataset}

To assess the performance of the proposed model, we utilized the public LIDC-IDRI dataset \cite{armato2011lung}. This dataset contains CT scans of nodules from 1,010 patients, which were identified and delineated by multiple radiologists and obtained by various medical centers in the United States using diverse imaging protocols and different scanner models from four manufacturers. We selected nodules that had a diameter greater than 3 mm and were identified and rated for attributes by at least one radiologist (from 875 patients), as done in previous studies \cite{liu2019multi}.

Nine high-level concepts were scored independently by up to 4 radiologists per nodule, meaning that not every nodule was evaluated by 4 radiologists. The attributes were rated on a scale of 1 to 5, except for calcification and internal structure concepts (they were evaluated as categories). For example, a score of 1 for the texture indicated that the nodule has the ground glass internal texture, a score of 5 for the margin indicated that the nodule has sharp margin. All used lung nodule concepts described in Table~\ref{tab:dataset_explanation}.

There is a variability between the number of rating instances per nodule and the scores of different radiologists. Therefore, we use the mean scores of the radiologists for each concept as the ground truth. The ground truth concept values are then normalized to be within the range of $[0, 1]$ for model training.

\begin{table}[!t]
\caption{The lung nodule concepts \& explanations}
\label{tab:dataset_explanation}
\begin{tabularx}{\columnwidth}{
     >{\raggedright\arraybackslash} p{0.2\columnwidth} X
     >{\raggedright\arraybackslash} p{0.4\columnwidth} X
     >{\raggedright\arraybackslash} X
}
\toprule
Feature type & Nodule concept & Concept rating system\\
\midrule
\multirow{7}{*}{Ordinal} 
& Subtlety & Extremely Subtle $\rightarrow$ Obvious\\ 
& Sphericity & Linear $\rightarrow$ Round \\ 
& Margin & Poorly Defined $\rightarrow$ Sharp \\ 
& Lobulation & No Lobulation $\rightarrow$ Marked \\ 
& Spiculation & No Spiculation $\rightarrow$ Marked \\ 
& Texture & GGO $\rightarrow$ Solid \\ 
& Malignancy & Unlikely $\rightarrow$ Suspicious \\ 
\midrule
\multirow{10}{*}{Categorical}
& \multirow{4}{*}{Internal structure} 
& Soft Tissue \\ 
& & Fluid \\ 
& & Fat \\ 
& & Air \\ 
\cmidrule{2-3}
& \multirow{6}{*}{Calcification} 
& Popcorn \\ 
& & Laminated \\ 
& & Solid \\ 
& & Non-central \\ 
& & Central \\ 
& & Absent \\ 

\bottomrule
\end{tabularx}
\end{table}

\subsubsection{Implementation details}

For the image processing, we utilize the pylidc python package, specifically designed for the LIDC-IDRI dataset, to extract the volume of interest (VOI) of each nodule \cite{hancock2016lung}. The VOI is determined by the common bounding box of the radiologists annotations, computed via a consensus consolidation at a 50\% agreement level. The VOIs were then adaptively padded to ensure they were cube-shaped. Only slices containing nodule-positive pixels, as indicated by the annotations, were taken into account.

All volumes are resized to the resolution $64 \times 64 \times 64$ using bilinear interpolation. The slice image intensities are then converted to Hounsfield units in a range of $[-1000, 700]$ and normalised to be within the range of $[0, 1]$. In total, there are 2651 nodules in the dataset.

We use ResNet-50 \cite{he2016deep}, DenseNet-121 \cite{huang2017densely} and SEResNet-50 \cite{hu2018squeeze} architectures as the proposed shared feature extractor $\phi$. To ensure a fair comparison, we maintain consistent implementation and hyperparameters throughout all experiments. The networks are trained from scratch for 80 epochs with a batch size of 16. We use the Adam optimizer with a dynamic learning rate from the base value 0.0001 and reduce the learning rate by factor 0.9 when a validation loss has stopped improving 4 epochs to minimize the total multi-task loss. The optimizer has the first moment estimate 0.9 and the second moment estimate 0.999.

We augment the training data using standard image data augmentation techniques. This involved randomly applying transformations such as flipping or random rotating the input volume in a range $[-15, 15]$ degrees, reversing the z-order of slices. We ensure that all slices in a nodule volume have the same transformation applied. We shuffle the order of training examples every epoch. Our models are built using Tensorflow and data processing pipeline is built with Tensorflow data API.

For each experimental setup, we apply the 5-fold cross-validation to the data. The nodules are randomly divided into two sets: a training set (80\% of the samples) and a testing set (20\% of the samples). This resulted in five folds with 2,121 training samples and 530 testing samples. We make sure that no nodule are included in both the training and testing sets of a fold. The obtained primary performance metric is the F1-score per categorical concept and the mean absolute error (MAE) per numerical concept, which measures the difference between the mean radiologist scores and the predictions made by each method. This metric is chosen because it has been widely used in previous studies.

\begin{table}
\caption{Concept prediction performance for different feature extractors}
\label{tab:predict_performance}

\begin{tabularx}{\columnwidth}{
     >{\raggedright\arraybackslash} p{0.25\columnwidth} X
     >{\raggedright\arraybackslash} X
     >{\raggedright\arraybackslash} X
     >{\raggedright\arraybackslash} X
}
\toprule
Concept prediction performance & ResNet-50 & DenseNet-121 & SEResNet-50\\
\midrule

Subtlety & 0.163 & \textbf{0.161} & 0.162\\
Sphericity & 0.146 & \textbf{0.139} & 0.141\\
Margin & 0.135 & 0.136 & \textbf{0.133}\\
Lobulation & 0.120 & 0.125 & \textbf{0.115}\\
Spiculation & 0.106 & 0.115 & \textbf{0.105}\\
Texture & 0.096 & \textbf{0.094} & 0.104\\
Malignancy & 0.131 & \textbf{0.114} & 0.115\\
\midrule
Internal structure & \textbf{0.990} & \textbf{0.990} & 0.982\\
Calcification & 0.964 & 0.962 & \textbf{0.967}\\

\bottomrule
\end{tabularx}
\end{table}

\subsection{Quantitative Evaluation}

In Table \ref{tab:predict_performance}, we present the prediction performance (the mean absolute error and the F1-score) per concept of our method and compare it for each adopted feature extractor architecture: ResNet-50, DenseNet-121, and SEResNet. 

According to the results presented in Table \ref{tab:predict_performance}, the model that utilizes DenseNet121 as its raw feature extractor outperforms in comparison to other CNN backbones, with an advantage of 5 concepts. The results suggest that detecting subtlety (difficulty of detection, with higher values indicating easier detection) is more challenging for all raw feature predictors, as indicated by a higher mean absolute error (MAE) compared to other concepts.

\subsection{Qualitative Evaluation}

\begin{figure*}
\centering
\includegraphics[width=1\linewidth]{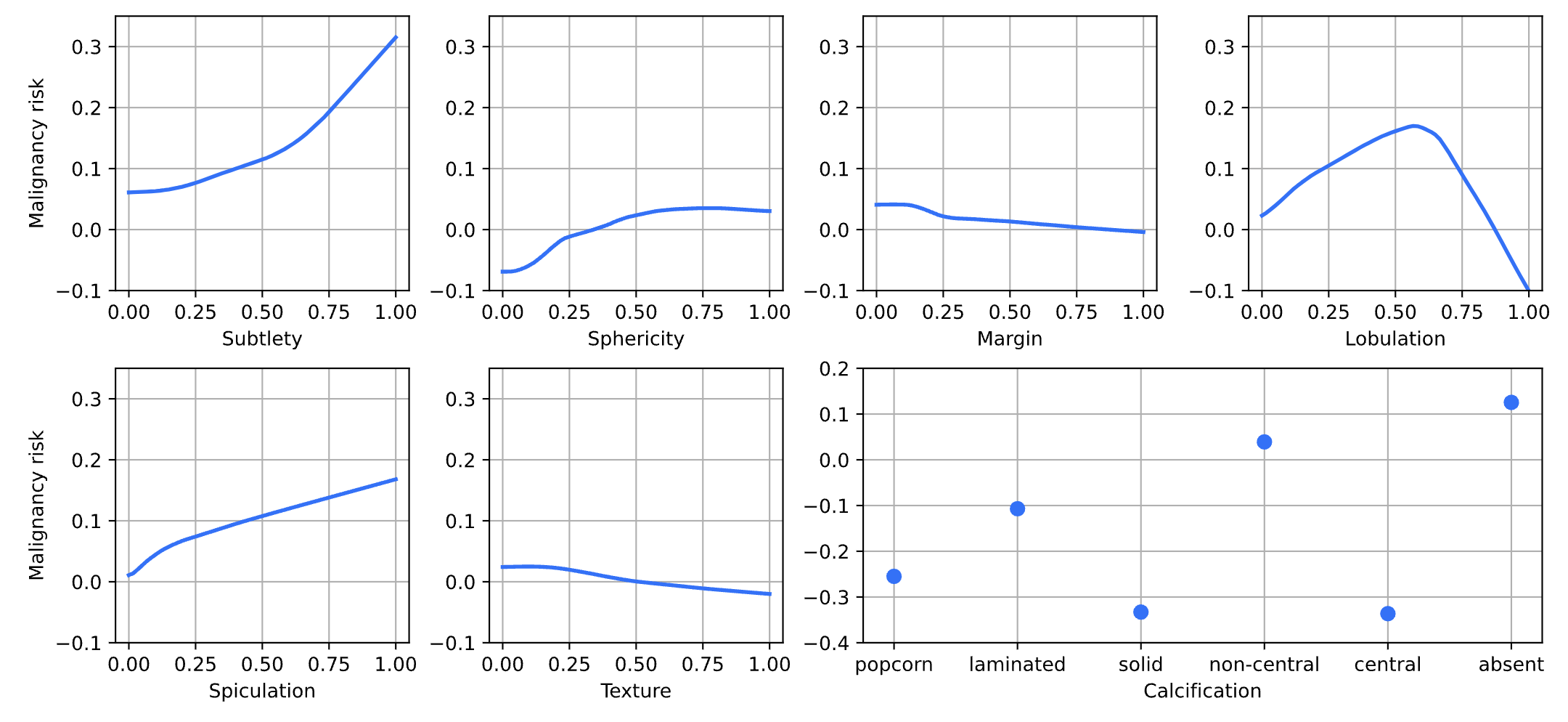}
\caption{\textbf{Concepts contributions to the malignancy risk.}  These plots show the individual shape functions learned by an Decision Explainer module using GAMs for each input concept.}
\label{fig:shape_functions}
\end{figure*}

To visualize how concept values affect malignancy scores at the model level rather than at the individual sample level, we depicted the trained shape functions in the decision explainer module for each concept in Figure \ref{fig:shape_functions}.

As we can see on Figure \ref{fig:shape_functions} the absence of calcification makes some positive contribution to the assessment of a nodule as malignant. It is confirmed in lung cancer screening that calcification in a nodule is usually associated with benign conditions \cite{snoeckx2018evaluation}. On the other side, high value (solid) of nodules texture makes negative contribution to the malignancy score due to part-solid nodules have a significantly higher risk of malignancy compared to solid nodules \cite{yip2016lung}. Also we can see that higher spiculation value has more contribution to the malignancy score compared to lower values of spiculation. This is also confirmed by the fact that the nodule with a spicular margin is much more likely to be malignant than one with a well-defined and smooth edge \cite{seemann1999usefulness, gurney1993determining}.

The attributes that posed the most challenges for our model based on any raw feature extractor architecture were sphericity, subtlety and margin, as shown in Table \ref{tab:predict_performance}. This was in line with the varying degrees of difficulty indicated by the radiologists annotations. The significant disparities among the radiologists annotations scores a high degree of ambiguity in the labeling process. For example, some nodules that appeared noticeably bright were rated as "Moderately Subtle" whereas nodules that were equally noticeable were rated differently.

\begin{table}
\caption{ CT image slice of three nodules along with predictions from the proposed model}
\label{tab:predictions_samples}

\begin{tabularx}{\columnwidth}{
     >{\raggedright\arraybackslash} p{0.22\columnwidth} X
     >{\raggedright\arraybackslash} X
     >{\raggedright\arraybackslash} X
     >{\raggedright\arraybackslash} X
     >{\raggedright\arraybackslash} X
     >{\raggedright\arraybackslash} X
     >{\raggedright\arraybackslash} X
}
\toprule
{} 
    & \multicolumn{2}{l}{\includegraphics[width=0.1\textwidth]{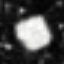}} 
    & \multicolumn{2}{l}{\includegraphics[width=0.1\textwidth]{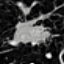}}
    & \multicolumn{2}{l}{\includegraphics[width=0.1\textwidth]{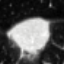}}
\\
\midrule
{} & PD & GT & PD & GT & PD & GT \\
Subtlety & 0.96 & 1.00 & 0.93 & 1.00 & 0.98 & 1.00\\
Sphericity & 0.77 & 0.81 & 0.59 & 0.75 & 0.71 & 0.75\\
Margin & 0.80 & 1.00 & 0.51 & 0.50 & 0.81 & 0.88\\
Lobulation & 0.36 & 0.38 & 0.48 & 0.56 & 0.47 & 0.12\\
Spiculation & 0.28 & 0.12 & 0.66 & 1.00 & 0.23 & 0.38\\
Texture & 0.98 & 1.00 & 0.98 & 0.94 & 0.99 & 1.00 \\
\midrule

Malignancy & 0.76 & 0.69 & 0.86 & 0.94 & 0.76 & 0.75 \\
\midrule

Internal structure & \multicolumn{2}{c}{Soft Tissue} & \multicolumn{2}{c}{Soft Tissue} & \multicolumn{2}{c}{Soft Tissue}\\
Calcification & \multicolumn{2}{c}{Absent} & \multicolumn{2}{c}{Absent} & \multicolumn{2}{c}{Absent}\\
\bottomrule
\end{tabularx}
\end{table}

\begin{figure*}
\centering
\includegraphics[width=1\linewidth]{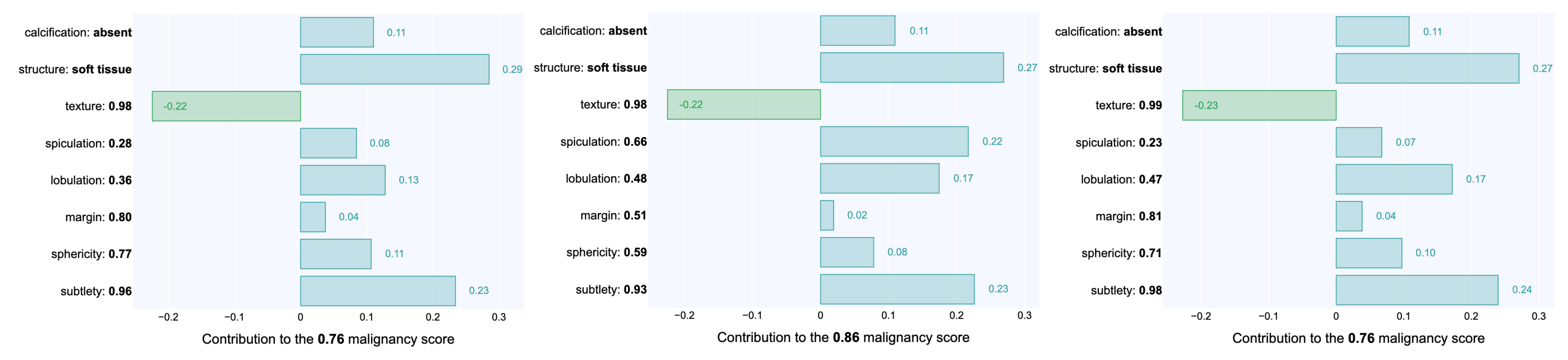}
\caption{\textbf{Predicted concepts contributions of three nodules.} Examples of provided concept contributions to the predicted malignancy score by our method for three nodules from Table \ref{tab:predictions_samples}.}
\label{fig:contributions}
\end{figure*}

We have incorporated some sample predictions generated by our model in Table \ref{tab:predictions_samples} for illustrative purposes, where GT is ground truth and PD is prediction. Their related predicted concept contributions to the malignancy score by our method presented in Figure \ref{fig:contributions}.

\section{Conclusion}

We have developed a model that can make decisions and provide explanations for those decisions using concepts that humans can understand. Our model's interpretability is achieved by calculating the contribution of each concept to the final decision. To do this, we have added a concept predictor component to a raw shared feature extractor. The concept predictor identifies the presence of a particular concept within an image or predicts the score of a numerical concept.

Previous research has focused on the indirect relationship between malignancy and other attributes. Malignancy or benignity can be inferred by certain concepts, such as patterns of calcification, the definition of the nodule margin, or the internal composition of the nodule. Previous studies have used a two-step approach, first predicting these characteristics and then using intermediate CNN features to infer malignancy without calculating the contribution values of concepts to the final malignancy regression score. We have added generalized additive sub-models to our framework, which allows us to disentangle the decision-making process and better understand the relationship between different concepts and the final decision. By using generalized additive sub-models, we can also decompose the decision-making process into additive components, making it easier to interpret and explain the model's predictions.

\bibliographystyle{IEEEtran}
\bibliography{IEEEabrv, main}

\end{document}